\documentclass[a4paper,11pt]{article}
\usepackage{pos}
\usepackage{siunitx}
\DeclareSIUnit\waterequivalent{w.e.}

\title{A New Method for Measuring the Pion-Air Cross Section at Multi-TeV Energies Using Muon Bundle Properties in Deep Underground Detectors}
\ShortTitle{A New Method for Measuring the Pion-Air Cross Section}

\author*[a]{Karolin Hymon}
\author[a]{Anatoli Fedynitch}

\affiliation[a]{Institute of Physics, Academia Sinica, Taipei City, 11529, Taiwan}

\emailAdd{khymon@as.edu.tw}
\emailAdd{anatoli@as.edu.tw}

\abstract{The interaction cross section of charged pions with air nuclei is a critical parameter for accurately simulating extensive air showers. Improving the modeling of high-energy pion interactions is essential for addressing the muon puzzle—the observed deficit of muons in simulations compared to indirect experimental estimates. As collider experiments cannot directly probe these interactions, we propose a novel measurement approach using muon bundles detected in deep-underground water Cherenkov detectors, such as IceCube and KM3NeT. This method aims to constrain the pion–air inelastic cross section, thereby reducing uncertainties in air shower simulations and advancing our understanding of cosmic ray interactions.
}

\ConferenceLogo{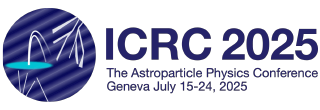}

\FullConference{39th International Cosmic Ray Conference (ICRC2025)\\
 15–24 July 2025\\
Geneva, Switzerland\\}

\begin{document}
\maketitle

\section{Introduction}

When a cosmic ray (CR) interacts with an atmospheric nucleus—typically nitrogen or oxygen—it initiates an extensive air shower, a cascade of secondary particles. The development of the muonic and neutrino components within this cascade is primarily governed by hadronic interactions between secondary particles and atmospheric nuclei, occurring in a kinematic regime inaccessible to current accelerator-based experiments. Accurate modeling of these hadronic interactions is crucial for reliable predictions of atmospheric muon and neutrino fluxes, which are essential for interpreting data from underground particle detectors, such as neutrino telescopes, and for interdisciplinary applications like muon tomography.

Despite recent advancements in quantifying uncertainties associated with these flux models \cite{Fedynitch:2021ima,Fedynitch:2022vty,Yanez:2023lsy,Woodley:2024eln}, significant challenges persist. In inclusive flux modeling, the primary challenge lies in the degeneracy between uncertainties arising from the cosmic-ray flux composition and those intrinsic to hadronic interaction models, predominantly involving nucleon-air interactions. In contrast, detailed studies of individual air showers explore a distinct regime dominated by pion-air interactions. In both contexts, notable discrepancies remain between contemporary simulations and observational data, with simulations systematically underestimating observed muon fluxes or counts by up to several tens of percent \cite{Riehn:2024prp,Albrecht:2021cxw}.

A key contributor to these discrepancies is the limited experimental constraints on pion-induced interactions. Accelerator-based measurements have currently only probed energies up to a few hundred GeV in the laboratory frame, most recently through the NA61 experiment \cite{NA61SHINE:2022tiz}. Even these measurements exhibit limited coverage in forward phase-space regions. At energies in the TeV range and beyond, no experimental constraints on fundamental quantities such as the inelastic pion-air cross section $\sigma_{\pi\mathrm{+air}}$ exist. Given the foreseeable difficulty of conducting high-energy pion-nucleon collision experiments at accelerator facilities, we propose here an alternative approach, leveraging indirect constraints obtained from cosmic-ray measurements. Demonstrations of the capability of air shower experiments to provide significant constraints on hadronic cross sections include measurements by the Pierre Auger Observatory and the Telescope Array at ultra-high energies \cite{PierreAuger:2012egl,Abbasi:2020chd}.

At the critical energy, the interaction and decay lengths of atmospheric hadrons become comparable. For pions interacting in air, this occurs at energies of approximately \SIrange{100}{200}{\giga\electronvolt}. Above this critical energy, an increase in the inelastic cross section enhances interaction probabilities, suppressing higher-energy secondary particle fluxes and consequently boosting the yields of lower-energy tertiary particles. This mechanism modifies the muon multiplicity distribution, establishing a direct connection between observed muon multiplicities underground and the inelastic pion-air cross section.

In parallel, the proton-air cross section $\sigma_{\mathrm{p+air}}$ governs the altitude at which CR primaries  initiate the air shower cascade. Increasing $\sigma_{\mathrm{p+air}}$ shortens the interaction length, initiating the cascade at higher altitudes and thereby altering both the energy and longitudinal development of the shower. This modification affects not only the surface muon flux but also, through subsequent meson production and decay processes, the resulting underground muon multiplicity distribution. However, proton-air interactions benefit from tighter experimental constraints derived from accelerator measurements, notably recent proton-oxygen collision data obtained at the LHC, thus offering better control on modeling uncertainties.

\section{Calculation of Underground Muon Multiplicities} 
\label{sec:mudist}

To quantitatively assess the relation between the inelastic pion-air and proton-air cross section and the underground muon multiplicity distribution, we employ the iterative cascade equation solver MCEq \cite{Fedynitch:2015zma,Fedynitch:2018cbl}. Given an initial  CR energy spectrum at the top of the atmosphere, MCEq computes the resulting spectrum of secondary particles at ground level. The initial condition can be specified either as a broad energy spectrum (primary flux) or as single hadrons placed at specific energies $E_\text{CR}$ using \texttt{MCEqRun.set\_single\_primary\_particle}. The calculated secondary muon spectrum at ground, denoted by $\mathrm{d}N_\mu/\mathrm{d}E_\mu$, defines the muon yield $\mathcal{Y}(E_\text{CR})$. The mean muon multiplicity above an energy threshold $E_\text{threshold}$ is then determined by integrating this yield:

\begin{equation}
\langle N_\mu (E_\text{CR}) \rangle = \int_{E_\text{threshold}}^\infty \mathrm{d}E_\mu~\mathcal{Y}(E_\text{CR}).
\end{equation}

Convolving the mean muon multiplicity with a cosmic ray spectrum model $\Phi_\text{CR}(E_\text{CR})$, such as the Global Spline Fit (GSF) model \cite{Dembinski:2017zsh}, allows the computation of the muon multiplicity "flux" through a transformation on the $E_\text{CR}$ grid:

\begin{equation}
\frac{\mathrm{d}N_\mu}{\mathrm{d}E_{\mathrm{CR}}} = \sum \langle N_\mu (E_\text{CR}) \rangle \Phi_\text{CR}(E_\text{CR}).
\end{equation}

The inclusive muon bundle flux $\mathcal{M}$, in units of \si{\per\centi\metre\squared\per\second\per\steradian}, measurable by extended-duration experiments, is obtained by applying the transformation:

\begin{equation}
\mathcal{M}(N_\mu) := \frac{\mathrm{d}N}{\mathrm{d}N_{\mu}} = \Phi_\text{CR}(E_\text{CR}(N_\mu)) \frac{\mathrm{d}E_\text{CR}}{\mathrm{d} N_{\mu}},\label{eq:mmd}
\end{equation}
where the derivative $\mathrm{d}E_{\mathrm{CR}}/\mathrm{d}N_{\mu}$ is computed numerically, as the muon multiplicity is a monotonic function of the cosmic ray energy.

The MUTE v3 code \cite{Woodley:2024eln} facilitates the calculation of these variables at arbitrary underground depths, expressed in kilometers of water equivalent (\si{\kilo\meter\waterequivalent}). Extending the calculation underground is straightforward, as MUTE employs a fast convolution method to generalize muon yields from $\mathcal{Y}(E_\text{CR})$ to $\mathcal{Y}(E_\text{CR}, X)$ at any depth $X$. Technically, the yields also depend on the zenith angle $\Theta$, but for simplicity we concentrate on the vertical direction.

\subsection{Numerical Example}
\begin{figure}
    \centering
    \begin{minipage}[t]{0.48\linewidth}
        \centering
        \includegraphics[width=\linewidth]{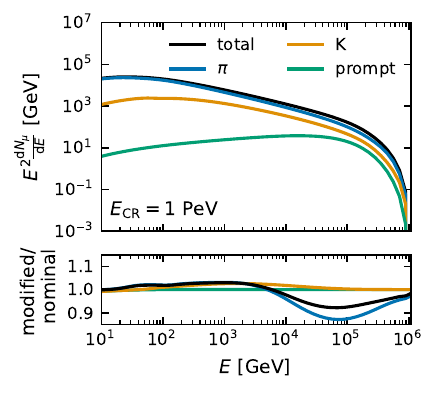}
        \caption{\textit{Top panel:} Vertical muon flux at the surface for a primary proton with an energy of \SI{1}{\peta\electronvolt}, computed at the South Pole using the SIBYLL~2.3c model. Relative contributions from different parent particle species are indicated. \textit{Bottom panel:} Ratio of the modified muon flux when increasing the pion-air cross section $\sigma_{\pi\mathrm{+air}}$ by 50\% in the $\sim$\SI{1}{\tera\electronvolt} region relative to the nominal cross section.}
        \label{fig:surf_spectrum}
    \end{minipage}
    \hspace{1em}
    \begin{minipage}[t]{0.48\linewidth}
        \centering
        \includegraphics[width=\linewidth]{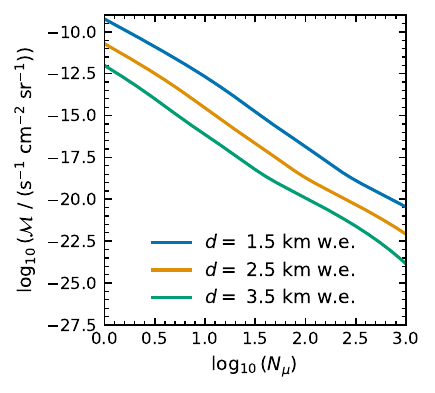}
        \caption{Vertical muon multiplicity distributions computed for proton primaries using MCEq and MUTE at several underground depths, assuming the GSF CR composition and the SIBYLL~2.3c hadronic interaction model.}
        \label{fig:dNdMu_p}
    \end{minipage}
\end{figure}

For the numerical example presented here, the simulation location is set to the South Pole site of the IceCube Neutrino Observatory \cite{IceCube:2016zyt}. Since all computations involve mean quantities, the superposition approximation for CR nuclei is valid, making it unnecessary to calculate muon yields separately for individual nuclear species. It is sufficient to simulate nucleons—in this case, protons—and utilize the nucleon CR flux $\Phi_\text{CR}(E_\text{CR})$ in the convolution. The hadronic interaction model employed is SIBYLL~2.3c \cite{Fedynitch:2018cbl,Riehn:2019jet}. The underground muon energy threshold is set to $E_\text{threshold} = \SI{100}{\giga\electronvolt}$, consistent with the energy threshold of the main detection array of IceCube.

Figure~\ref{fig:surf_spectrum} (top panel) illustrates the vertical muon flux at ground level resulting from a proton primary of \SI{1}{\peta\electronvolt}. The relative contributions from different parent particle species clearly  indicate that pions dominate muon production across the entire muon energy range. The bottom panel quantifies the effect of a hypothetical 50\% increase in the pion-air inelastic cross section around the \SI{1}{\tera\electronvolt} scale. Such a modification leads to a suppression of the muon flux at higher energies, coupled with an enhanced flux at lower energies. Corresponding effects from modifications in kaon-air interactions are significantly smaller and thus neglected in the current analysis.

Figure~\ref{fig:dNdMu_p} displays the resulting vertical muon multiplicity distributions at three distinct underground depths. These distributions exhibit a power-law-like behavior with subtle spectral features that reflect the structure of the primary CR flux.

\subsection{Modifying the Inelastic Cross Section}

To avoid model-specific assumptions, the inelastic cross section from SIBYLL~2.3c is replaced by a cubic spline interpolation, denoted as $\sigma_{\hat{c}_i}$, fitted to the original tabulated cross section. This spline is constructed using 10 knots, equally spaced in logarithmic energy from \SI{10}{\giga\electronvolt} to \SI{e10}{\giga\electronvolt}. The spline is fully defined by 12 coefficients $c_i$: 10 coefficients at the internal knots and 2 additional coefficients to serve as edge padding. We precompute a nominal set of muon yields $\mathcal{Y}(E_\text{CR})$, along with yields obtained by slightly perturbing each spline parameter to numerically estimate gradients via central finite differences:
\begin{equation}
\frac{\partial \mathcal{M}}{\partial \hat{c}_i} = \frac{\mathcal{M}\left((1+\delta)\hat{c}_i\right) - \mathcal{M}\left((1-\delta)\hat{c}_i\right)}{2\delta}. \label{eq:gradient}
\end{equation}

\noindent The modified muon multiplicity distribution incorporating these perturbations is then expressed as:
\begin{equation}
\mathcal{M}_{\text{model}} (N_{\mu}, d, c_1, \ldots, c_n) = \mathcal{M}_{\text{nominal}} + \sum_i \frac{\partial \mathcal{M}}{\partial \hat{c}_i} c_i. \label{eq:tot_mmodification}
\end{equation}
The tuning parameters to the spline coefficients $c_i$ are the parameters optimized in the subsequent sensitivity analysis.

To create pseudo-data, we adopt the nominal multiplicity distribution $\mathcal{M}_{\text{nominal}}$ and assume a conservative uncertainty of $\log_{10}(\Delta \mathcal{M}/\mathcal{M}) = 0.1$. This assumption is motivated by IceCube measurements \cite{IceCube:2025baz}, which demonstrate multiplicity reconstruction accuracies ranging from 5\% to 20\% and report billions of events extending to multiplicities exceeding $10^3$ muons per bundle. While precise event-by-event multiplicity reconstruction is not strictly required, accurate simulation of the detector response to muon bundles remains crucial.

Sensitivity to the cross section spline parameters is estimated by minimizing a penalized $\chi^2$ function, quantifying the differences between pseudo-data and model predictions:

\begin{equation}
\chi^2 = \sum_n \left(\frac{\log_{10}(\mathcal{M}^{\text{nominal}}) - \log_{10}(\mathcal{M}^{\text{model}}(\vec{c}))}{0.1}\right)^2 + \sum_i \left(\frac{c_i - c_i^{\text{prior}}}{\delta_{c_i}}\right)^2, \label{eq:chi2}
\end{equation}

\noindent Gaussian priors centered at the nominal cross section ($c_i^{\text{prior}} = 0$) are employed for regularization, with a broad prior width of $\delta_{c_i} = 5$ to ensure numerical stability while preventing unphysical distortions at the energy boundaries. The projected uncertainty on $\sigma_{\pi+\mathrm{air}}(E)$  and $\sigma_{\mathrm{p+air}}(E)$ is obtained by propagating the covariance of the fitted spline coefficients through first-order linear error propagation:

\begin{equation}
\mathrm{Var}[\sigma_{\pi+\mathrm{air}}(E)] = \mathbf{J}^T \mathbf{\Sigma}_c \mathbf{J},
\label{eq:cs_variance}
\end{equation}

\noindent where $\mathbf{J}$ is the Jacobian matrix containing the gradients $\partial \sigma / \partial c_i$ evaluated at each energy, and $\mathbf{\Sigma}_c$ represents the parameter covariance matrix derived from the inverse Hessian evaluated at the minimum of the $\chi^2$ function. Thus, the inferred uncertainties on the cross sections represent the anticipated sensitivity achievable given the assumed experimental precision in measuring muon multiplicity distributions.

\section{Impact of Cross Section Modifications on the Multiplicity Distribution}

\begin{figure}
    \centering
    \includegraphics[width=\textwidth]{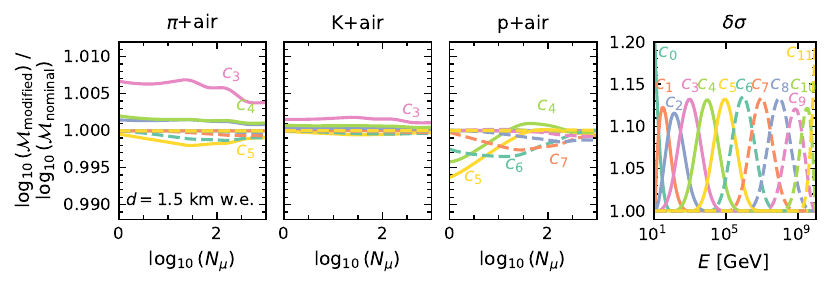}
    \caption{Illustration of the impact on the muon multiplicity distribution resulting from variations of the inelastic cross sections within specific energy ranges defined by spline coefficients, relative to the nominal SIBYLL~2.3c predictions. \textit{Left panels}: Effects on the multiplicity distribution at a depth of $d = \SI{1.5}{\kilo\metre\waterequivalent}$ induced by a 20\% increase ($c_i = + 0.2$) in the pion-, kaon-, and proton-air cross sections. \textit{Right panel}: Corresponding cross section modifications showing the energy range associated with each spline coefficient.}
    \label{fig:ratio_dNdNmu}
\end{figure}

Figure~\ref{fig:ratio_dNdNmu} demonstrates how the muon multiplicity distribution at a vertical depth of \SI{1.5}{\kilo\metre\waterequivalent} responds to individual variations of $+20$\% in spline coefficients for $\sigma_{\pi+\mathrm{air}}$, $\sigma_{\mathrm{K+air}}$ and $\sigma_{\mathrm{p+air}}$ as detailed in the preceding section. The energy ranges associated with each spline coefficient are depicted in the right panel. These results emphasize how localized modifications of the inelastic cross sections for $\pi$-air, K-air, and p-air interactions propagate through the atmospheric cascade and impact underground muon multiplicities.

$\mathcal{M}$ shows the greatest sensitivity to changes in $\sigma_{\pi+\mathrm{air}}$ at energies around a few \si{\tera\electronvolt}. In this regime, an increased cross section results in a notable rise in the overall underground muon multiplicity, highlighting the dominant role of lower-energy muon pile-up relative to the flux suppression at higher energies (as illustrated in Fig.~\ref{fig:surf_spectrum}). Variations in  $\sigma_{\mathrm{K+air}}$ produce significantly smaller effects, justifying their exclusion from detailed sensitivity analyses. Conversely, modifications to  $\sigma_{\mathrm{p+air}}$ exhibit a comparable magnitude to those from pion-air interactions but yield an opposite trend, decreasing rather than increasing the underground multiplicity.

\section{Projected Sensitivity to the Inelastic Pion-Air and Proton-Air Cross Sections}

\begin{figure}[tbh]
    \centering
    \includegraphics[width=0.99\linewidth]{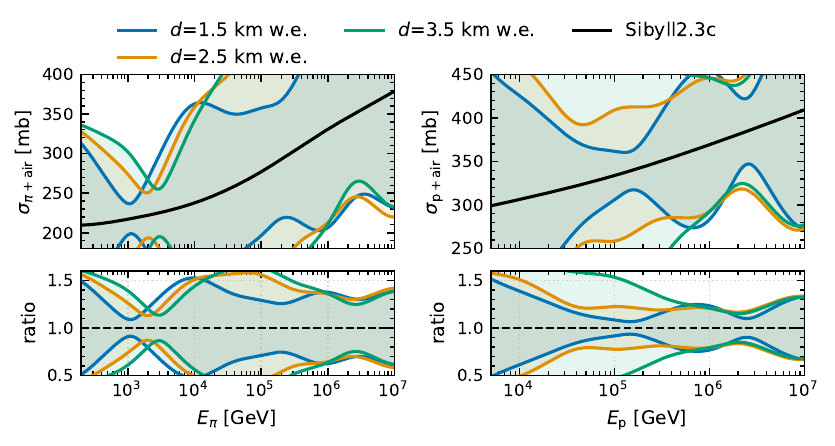}
    \caption{Projected sensitivities to $\sigma_{\pi+\mathrm{air}}$ and $\sigma_{\mathrm{p+air}}$ at various underground depths. The shift in the energy range of maximum sensitivity with increasing depth is clearly evident.}
    \label{fig:sensitivity_depth}
\end{figure}

The projected sensitivities to $\sigma_{\pi+\mathrm{air}}$ and $\sigma_{\mathrm{p+air}}$, calculated by Eq.~\ref{eq:cs_variance}, are presented in Fig.~\ref{fig:sensitivity_depth} for three vertical depths ranging from \SI{1.5}{\kilo\metre\waterequivalent} to \SI{3.5}{\kilo\metre\waterequivalent}.

For $\sigma_{\pi+\mathrm{air}}$, the projected sensitivity at the IceCube location peaks at approximately 10–15\% around \SI{700}{\giga\electronvolt} to \SI{2}{\tera\electronvolt}, with uncertainties below 50\% between \SI{200}{\giga\electronvolt} to \SI{6}{\tera\electronvolt} at a depth of \SI{1.5}{\kilo\metre\waterequivalent}. Additionally, a second region of enhanced sensitivity below 40\% emerges between \SI{30}{\tera\electronvolt} and \SI{10}{\peta\electronvolt}, peaking at~25–30\% around \SI{200}{\tera\electronvolt}. The energy corresponding to peak sensitivity shifts upwards by approximately \SI{1}{\tera\electronvolt} per \SI{1}{\kilo\metre\waterequivalent} of increased depth, reflecting the survival of higher-energy muons at greater depths. This higher-energy sensitivity region broadens progressively with increasing depth.

For $\sigma_{\mathrm{p+air}}$, the projected sensitivity at \SI{1.5}{\kilo\metre\waterequivalent} peaks at 10\% between \SI{80}{\tera\electronvolt} to \SI{200}{\tera\electronvolt}, and remains below 30\% from \SI{20}{\tera\electronvolt} to \SI{10}{\peta\electronvolt}. Muon multiplicity measurements demonstrate heightened sensitivity to $\sigma_{\mathrm{p+air}}$, maintaining lower uncertainties across nearly three orders of magnitude in energy. With increasing depth, the energy of peak sensitivity for proton-air interactions shifts upward by approximately \SIrange{5}{10}{\tera\electronvolt} at \SI{2.5}{\kilo\metre\waterequivalent} and by \SI{100}{\tera\electronvolt} at \SI{3.5}{\kilo\metre\waterequivalent}. At any fixed depth, the sensitivities for pion-air and proton-air interactions clearly occupy complementary, distinct energy ranges, as supported by the results shown in Fig.~\ref{fig:ratio_dNdNmu}.

\section{Conclusion}

We have introduced an indirect method to measure the inelastic pion–air and proton–air cross sections using underground muon multiplicity distributions. By employing a spline-based parametrization, we systematically varied these cross sections to examine the sensitivity of the muon multiplicities as an observable of these cross sections. Assuming an experimental uncertainty on the multiplicity of $\mathcal{O}(0.1)$ in $\log_{10}(N_{\mu})$ per bin, typical in underground neutrino telescopes such as IceCube, we propagated this uncertainty to estimate achievable constraints. Under these assumptions, the inelastic pion–air cross section can be constrained to approximately 10-15\% uncertainty from \SIrange{1}{2}{\tera\electronvolt}, while sensitivity to the proton–air cross section reaches similar precision around few \SI{100}{\tera\electronvolt}. These sensitivity regions partially overlap near \SI{100}{\tera\electronvolt}, but the distinct energy dependence and systematic depth dependence allow clear discrimination. Moreover, existing tighter laboratory constraints on the proton cross section are expected to further alleviate potential degeneracies.

Beyond this initial proof-of-principle demonstration, additional constraints can be achieved by incorporating measurements at multiple zenith angles. This extension would allow continuous probing of muon multiplicities across a range of effective depths, potentially broadening the sensitivity window from a few TeV into the multi-hundred TeV regime, with detailed energy segmentation yet to be determined. Given the limited current knowledge of pion-air and proton-air interactions at multi-TeV energies, this approach offers significant potential for reducing uncertainties in hadronic interaction modeling, with wide-ranging implications for cosmic-ray and neutrino physics.

\section*{Acknowledgments}
We acknowledge support from Academia Sinica (Grant No.~AS-GCS-113-M04) and the National Science and Technology Council (Grant No.~113-2112-M-001-060-MY3). K.~Hymon acknowledges support from the Postdoctoral Scholar Program of the Academia Sinica (PD-11401-M-3854).
\let\oldbibliography\thebibliography
\renewcommand{\thebibliography}[1]{%
  \oldbibliography{#1}%
  \setlength{\itemsep}{3pt}%
}

{\small
\bibliographystyle{JHEP}
\bibliography{references}
}

\end{document}